\def\year{2022}\relax
\documentclass[letterpaper]{article} 
\usepackage{aaai22}  
\usepackage{times}  
\usepackage{helvet}  
\usepackage{courier}  
\usepackage[hyphens]{url}  
\usepackage{graphicx} 
\usepackage{natbib}  
\usepackage{caption} 
\usepackage{subcaption}

\DeclareCaptionStyle{ruled}{labelfont=normalfont,labelsep=colon,strut=off} 
\usepackage{algorithm}
\usepackage{algorithmic}

\usepackage{newfloat}
\usepackage{listings}
\floatstyle{ruled}
\newfloat{listing}{tb}{lst}{}
\floatname{listing}{Listing}
\title{GeoCovaxTweets: COVID-19 Vaccines and Vaccination-specific Global Geotagged Twitter Conversations}
\author{
    Pardeep Singh$^{1,p}$, Rabindra Lamsal$^{2,r}$, Monika$^{1,m}$, Satish Chand$^{1,s}$, Bhawna Shishodia$^{1,b}$\\
}
\affiliations {
    \textsuperscript{\rm 1}School of Computer \& Systems Sciences, Jawaharlal Nehru University, New Delhi, India\\
    \textsuperscript{\rm 2} School of Computing and Information Systems, The University of Melbourne, Victoria, Australia\\

    \{$^p$pardee87\_scs@,$^m$monika95\_scs@,$^s$schand@mail.\}jnu.ac.in, $^r$r.lamsal@unimelb.edu.au, $^b$bhawna.shishodia.du@gmail.com
}

\usepackage{bibentry}

\begin{document}
 \nocopyright
\maketitle 

\begin{abstract}
Social media platforms provide actionable information during crises and pandemic outbreaks. The COVID-19 pandemic has imposed a chronic public health crisis worldwide, with experts considering vaccines as the ultimate prevention to achieve herd immunity against the virus. A proportion of people may turn to social media platforms to oppose vaccines and vaccination, hindering government efforts to eradicate the virus. This paper presents the COVID-19 vaccines and vaccination-specific global geotagged tweets dataset, \textbf{GeoCovaxTweets}\footnote{\texttt{https://doi.org/10.7910/DVN/YTH2MM}}, that contains more than 1.8  million tweets, with location information and longer temporal coverage, originating from 233 countries and territories between January 2020 and November 2022. The paper discusses the dataset's curation method and how it can be re-created locally, and later explores the dataset through multiple tweets distributions and briefly discusses its potential use cases. We anticipate that the dataset will assist the researchers in the crisis computing domain to explore the conversational dynamics of COVID-19 vaccines and vaccination Twitter discourse through numerous spatial and temporal dimensions concerning trends, shifts in opinions, misinformation, and anti-vaccination campaigns.
\end{abstract}

\section{Introduction} The COVID-19 outbreak in December 2019, resulted in more than six million deaths cases and 600 million confirmed cases globally \cite{Worldometer}. The respiratory illness caused by severe acute respiratory syndrome coronavirus 2 (SARS-CoV-2) was declared a public health emergency of international concern on January 30, 2020, and a pandemic on March 13, 2020, by World Health Organization. Vaccines such as Pfizer, Oxford-AstraZeneca, Johnson and Johnson, and Moderna were started in late 2020 to control the spread \cite{WHO}. To acquire herd immunity to end the pandemic, it has been estimated that around 60-70\% of the world population must get vaccinated \cite{aguas2022herd}. However, one of the biggest hindrances to vaccinations is hesitancy in acceptance due to numerous reasons, importantly perceived fears.

Myths and fears around vaccines are not new. The COVID-19 pandemic has brought vaccine hesitancy back to the limelight. Social media, which provides real-time access to people's opinions and beliefs across demographics, has frequently been signified as a hotbed of activity for anti-vaccination activists \cite{kata2012anti}. These activists may turn to social media platforms to oppose vaccinations with claims lacking scientific support. Metaphors used to argue against vaccination revolve around negative phrases such as ``No Forced Vaccines" and ``say no to vaccines".

\cite{KFF} reported that during February and March 2021, 40–47\% of American adults were hesitant towards COVID-19 vaccinations. Previous studies have linked social media and the anti-vaccination movement towards vaccine hesitancy \cite{johnson2020online,burki2019vaccine}. The COVID-19 vaccines and vaccination-specific discussions on social media platforms have influenced people's decisions to accept or reject vaccination. Therefore, understanding the nature of vaccine hesitancy through publicly available social media discussions for different geographical regions can open interesting research avenues in the crisis computing domain. Numerous COVID-19-specific datasets have been released for studying the nature of COVID-19 vaccine hesitancy.

\cite{chen2020tracking,lamsal2021design,banda2021large,imran2022tbcov} are some of the pubicly available large-scale COVID-19 datasets that have hundreds of millions of tweets. However, those datasets vaguely focus on general COVID-19 discussions as their keyword sets are not entirely focused on vaccines and vaccination-specific discourse. Closely focused datasets have also been introduced \cite{deverna2021covaxxy,muric2021covid,malagoli2021look}; however, such datasets are either focused on a particular geographical region or have limited temporal coverage. To bridge these gaps, we present a large-scale geotagged tweets dataset, \textit{GeoCovaxTweets}, containing vaccines and vaccination-related Twitter conversations with geolocation data. This dataset covers various perspectives on COVID-19 vaccines and vaccination, as our keywords and hashtags set captures a comprehensive conversational dynamics of support, criticism, and hesitance towards the COVID-19 vaccines and vaccination.

\section{Related Work}
Several researchers collect and share large-scale COVID-19 tweets datasets. \cite{chen2020tracking,banda2021large,lamsal2021design,qazi2020geocov19} provide large-scale tweet collections with different data distributions, collection periods, and sets of keywords. \cite{chen2020tracking,banda2021large} present multilingual datasets with 2.5 billion and 1.12 billion tweets respectively. Similarly, \cite{imran2022tbcov} has 2 billion multilingual tweets with the last release on March 2021. \cite{lamsal2021design} maintains an English-only collection with more than 2.2 billion tweets. The keywords sets differ amongst the datasets---\cite{chen2020tracking} use 80 keywords, \cite{banda2021large} use 10 keywords, \cite{lamsal2021design} uses 90+ keywords, \cite{imran2022tbcov} use 800+ multilingual keywords. The data collection of \cite{chen2020tracking,banda2021large,lamsal2021design} is still ongoing. The common issue with these billion-scale datasets is that they are designed to capture the conversational dynamics of COVID-19 across every thematic area. In doing so, due to limits (per day upper limit and sample size) set by Twitter on its filtered stream endpoint, these datasets contain proportionately fewer theme-specific tweets compared to theme-dedicated collections.

Vaccines and vaccination-focused datasets have also been released. \cite{muric2021covid} presented a dataset of 137 million tweets having antivaccine discussions. The dataset includes tweets fetched through two complementary collections: first, using 60 keywords through streaming API, and second, collecting historical tweets from 70k accounts that allegedly propagated anti-vaccine tweets. \cite{deverna2021covaxxy} introduced a dataset of 4.7 million tweets that includes one week of vaccine-related tweets and also designed a dashboard to track and quantify credible information and misinformation records along with their sources and keywords. Another paper by \cite{hu2021revealing} presented spatio-temporal trends from March 01, 2022, to February 08, 2022, toward COVID-19 vaccines in the United States. Similarly, vaccine hesitancy has been studied for Australia \cite{kwok2021tweet} and the globe \cite{chang2021people}.

However, the duration of these studies is either temporally-limited or mainly focuses on the pandemic's early stages. Hence, there is a need for critical work to explore the evolving nature of the public's perception of COVID-19 vaccines over an extended period. Additionally, most studies about COVID-19 vaccines are country specific with a significant focus on the United States.

To address the above gaps, we present a large-scale COVID-19 vaccines and vaccination-related geotagged tweets dataset with a cross-national, extended timeline, and pro-anti-hesitant focus. We used Twitter's Full-archive search endpoint to search and collect historical tweets; this endpoint returns 100\% of Twitter data for a \texttt{query+condition}, unlike filtered and standard search endpoints, used by the majority of the studies discussed above, which return 1\% of entire Twitter data at a particular time. We envision that the dataset will help the researchers in the crisis computing domain to analyze Twitter conversations for exploring the global spatio-temporal shifts and trends related to the COVID-19 vaccine narrative. We discuss potential use cases of the dataset later in the paper.

\section{Data Collection}
\begin{figure*}
\centering
\includegraphics[width=0.82\textwidth]{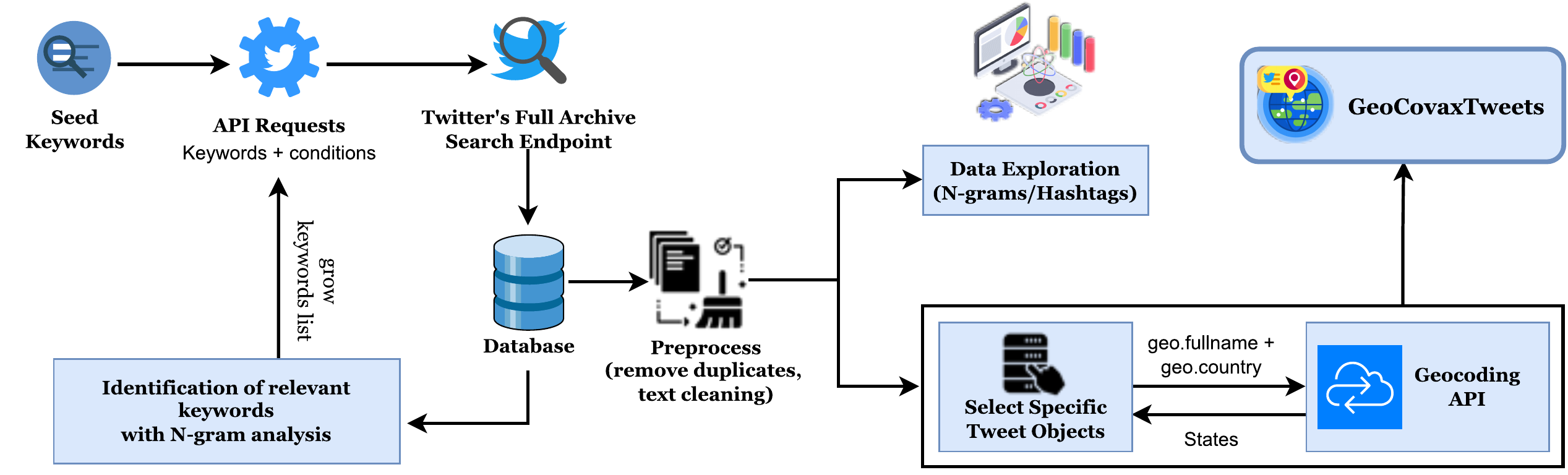}
    \caption{The data curation process.}
    \label{data-curation-process}
\end{figure*}

We used the \textit{twarc} python library to utilize Twitter's Full-archive search endpoint. The endpoint returns all historical tweets, unlike the standard search endpoint which returns 1\% of tweets at a particular time. To collect tweets relevant to COVID-19 vaccines and vaccination, we used meta-seed keywords having pro and anti-vaccine stances through a snowball sampling technique \cite{yang2019bot}. We started with the following seed keywords: \textit{CovidVaccineScam}, \textit{NoForcedVaccines}, \textit{covidvaccineVictims}, \textit{BanCovidVaccine}, \textit{getvaccinated}, \textit{covishield affect}, \textit{Covaxin affect}, \textit{vaccineinjuries}, \textit{NoForcedVaccination}, \textit{vaccine}, and \textit{vaccination}. Moving on, we analyzed $N$-grams in the collected corpus every 10 minutes to track emerging relevant keywords, similar to the strategy applied in \cite{lamsal2021design}. Furthermore, we refined the keyword list by identifying potential keywords that frequently co-occur with the seed keywords \cite{deverna2021covaxxy}. We provide the complete set of keywords and hashtags used for curating \textit{GeoCovaxTweets} in Table \ref{keywords-table}. Furthermore, we applied \textit{has:geo} and \textit{lang:en} conditional operators to collect only English-language tweets that are geotagged. The complete data curation process is illustrated in Figure \ref{data-curation-process}.

For each collected tweet, we also provide state information besides country information. Although the \textit{geo.full\_name} tweet object contains state information, some tweets had [city, state] or [state, country] information---for instance, [Mumbai, Maharastra], and [Maharastra, India]---and therefore normalization was required for consistency. We set up a planet-level geocoding endpoint powered by \textit{OpenStreetMap}\footnote{https://www.openstreetmap.org/} data. The endpoint was backed with a search index of 61 gigabytes\footnote{https://download1.graphhopper.com/public/} (when compressed) on a virtual machine with 24 VCPUs and 216 gigabytes of memory. The endpoint returned state information for each [city, state] or [state, country] request.

\begin{table*}
    \centering
    \caption{Keywords and hashtags used in curating \textit{GeoCovaxTweets}.}
    \begin{tabular}{p{\textwidth}}
\hline
vaccine, vaccines, vaccination, vaccinations, \#vaccine, \#vaccines, \#vaccination, \#vaccinations, \#coronavaccine, \#CovidVaccineScam, \#covidvaccinedeaths, \#covidvaccineVictims, \#CovidVaccineKills, \#crimesagainsthumanity, \#BanCovidVaccine, \#vaxxdamage, \#CovidVaccinesKill, Antivaccine, ****vaccines, forcedvaccines, NoVaccineForMe, NoVaccine, NoForcedVaccines, vaccinefailure, vaccinefraud, vaccinedamage, vaccineharm, unvaccinated, saynotovaccines, notomandatoryvaccines, covidvaccineispoison, learntherisk, vaccineinjuries, vaccineinjury, NoForcedVaccines, vaccinefailure, depopulation, stopmandatoryvaccination, covaxin serious problems, covishield serious problems, COVAXIN new born calf serum, Luc Montagnier's claim, Luc Montagnier's vaccine statement, abolishbigpharma, notocoronavirusvaccines, educateb4uvax, getvaccinated, iwillgetvaccinated, \#Covaxin affect, \#COVISHIELD affect, \#pfizervaccine affect, \#PfizerKills, \#modernavaccineaffect, \#Sputnikvaccine affect, \#SputnikV unsafe, \#Astrazenica affect, \#oxfordvaccine affect, \#VaccineSaveLives, \#VACCINESWORK \\
\hline
    \end{tabular}
    \label{keywords-table}
\end{table*}

\section{Using the Dataset}
 The \textit{GeoCovaxTweets} dataset is publicly available from \textit{Harvard Dataverse} at this URL: \texttt{https://doi.org/10.7910/DVN/YTH2MM}. We release the following tweet objects in the dataset: \textit{id}, \textit{created\_at}, \textit{author.id}, \textit{author.location}, \textit{source}, \textit{entities.hashtags}, \textit{geo.country\_code}, \textit{author.public\_metrics.followers\_count}, and \textit{state}. Twitter's data re-distribution policy\footnote{https://developer.twitter.com/en/developer-terms/agreement-and-policy} restricts the sharing of raw Twitter data with third parties. Therefore, tweet identifiers (i.e., \textit{id} tweet object) need to be hydrated to re-create the dataset locally. Hydration of tweet identifiers refers to the process of fetching raw tweet data using Twitter's tweet lookup endpoint. The lookup endpoint has a limit of 900 requests per 15-minute window; therefore, \textit{GeoCovaxTweets} can be re-created locally within a few hours.

Tools such as Hydrator (desktop application) and twarc (Python library) can be used to hydrate the tweet identifiers present in \textit{GeoCovaxTweets}. In both use cases---hydrating all the tweet identifiers in \textit{GeoCovaxTweets} or a subset---tweet identifiers should be in a file (e.g., TXT/CSV) with each identifier on a different line without any header and quotes. With the Hydrator app, we need to link a Twitter account for authorization and load the file with tweet identifiers for starting hydration. In the case of  twarc, we can use the following command:

Command: \texttt{
twarc2 hydrate --consumer-key consumer-key-here --consumer-secret consumer-secret-here --access-token access-token-here --access-tokensecret access-token-secret-here ids-file.txt data.jsonl
}

The above command hydrates the tweet identifiers in \texttt{ids-file.txt} and saves the API responses (hydrated data) in  \texttt{data.jsonl}. Such hydrated data can be converted to CSV\footnote{https://github.com/DocNow/twarc-csv} for loading the data as \textit{pandas} dataframe for convenience.

\section{The \textit{GeoCovaxTweets} Dataset}

In this section, we briefly explore the \textit{GeoCovaxTweets} dataset, present its potential use cases and provide additional information.

\subsection{Data Description}
\begin{figure*}
\centering
\includegraphics[width=1\textwidth]{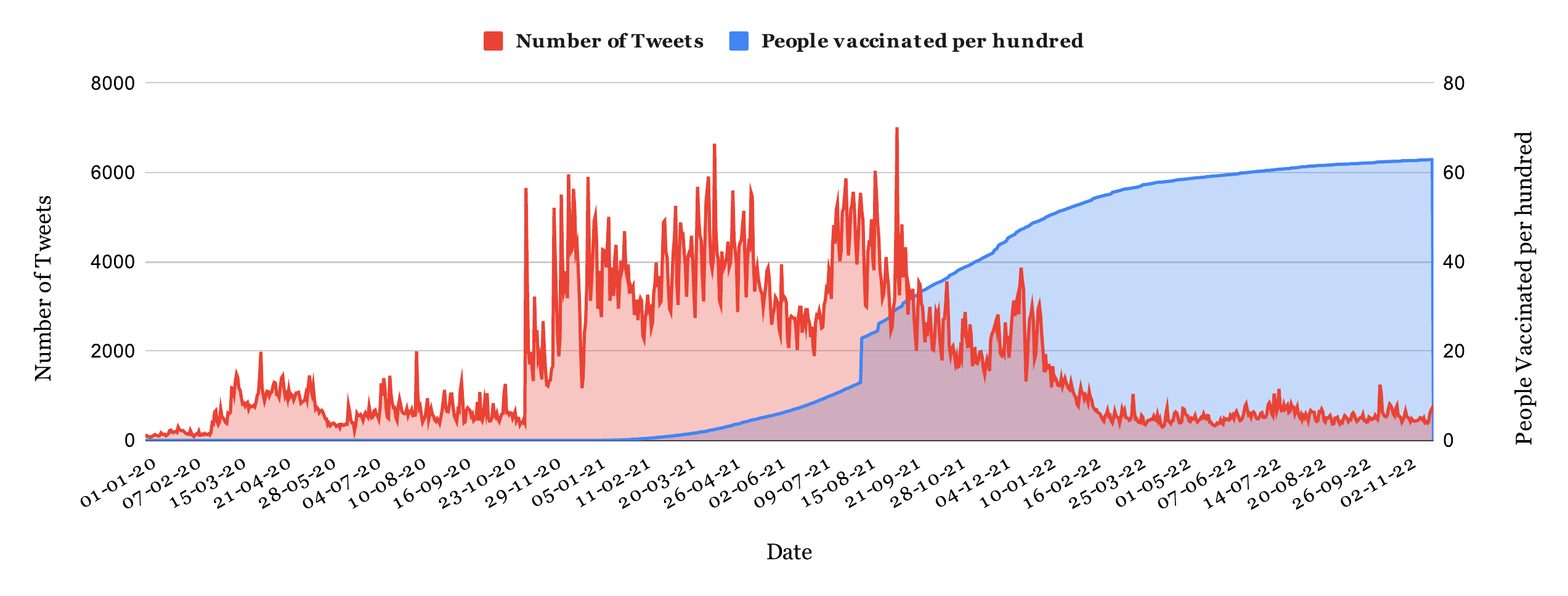}
\caption{The daily global distribution of the tweets and the cumulative number of people vaccinated per hundred.}
\label{daily-dist}
\end{figure*}

\begin{figure*}
     \centering
     \begin{subfigure}[b]{0.49\textwidth}
         \centering
         \includegraphics[width=0.86\textwidth]{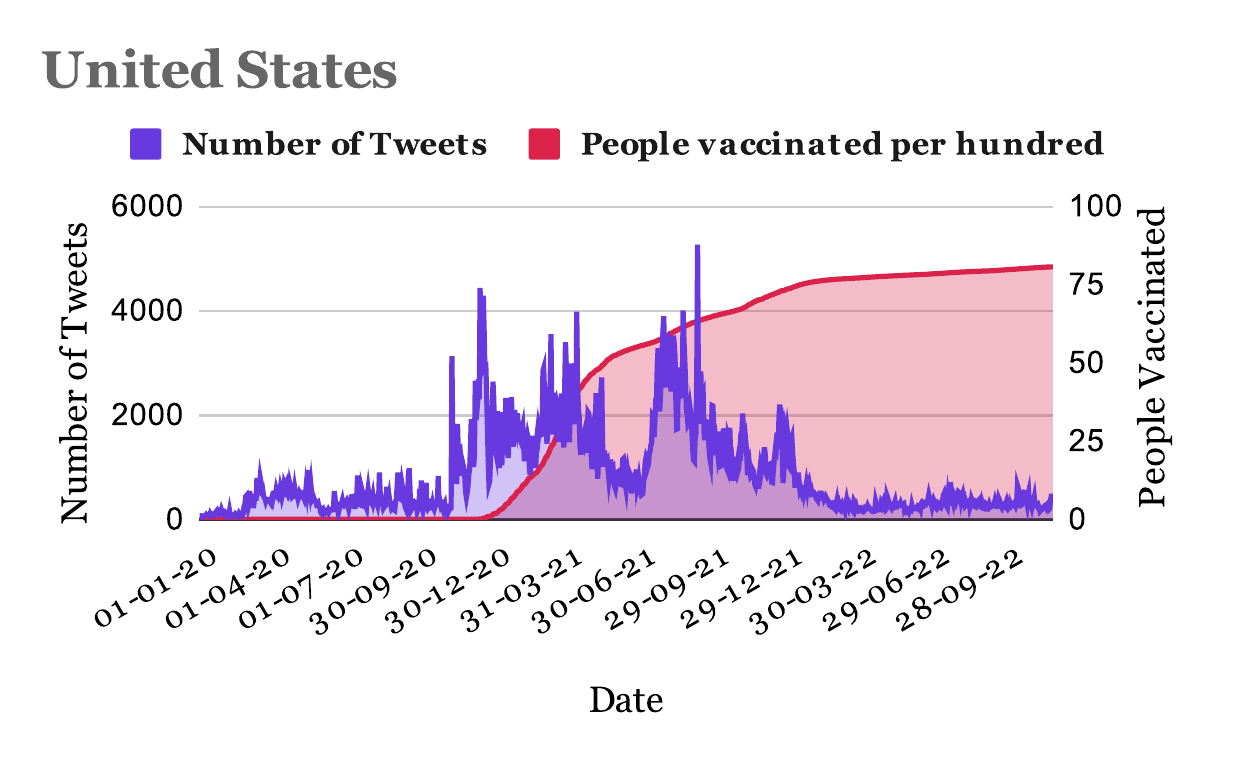}
         \label{fig:1}
     \end{subfigure}
     \hfill
     \begin{subfigure}[b]{0.49\textwidth}
         \centering
         \includegraphics[width=0.86\textwidth]{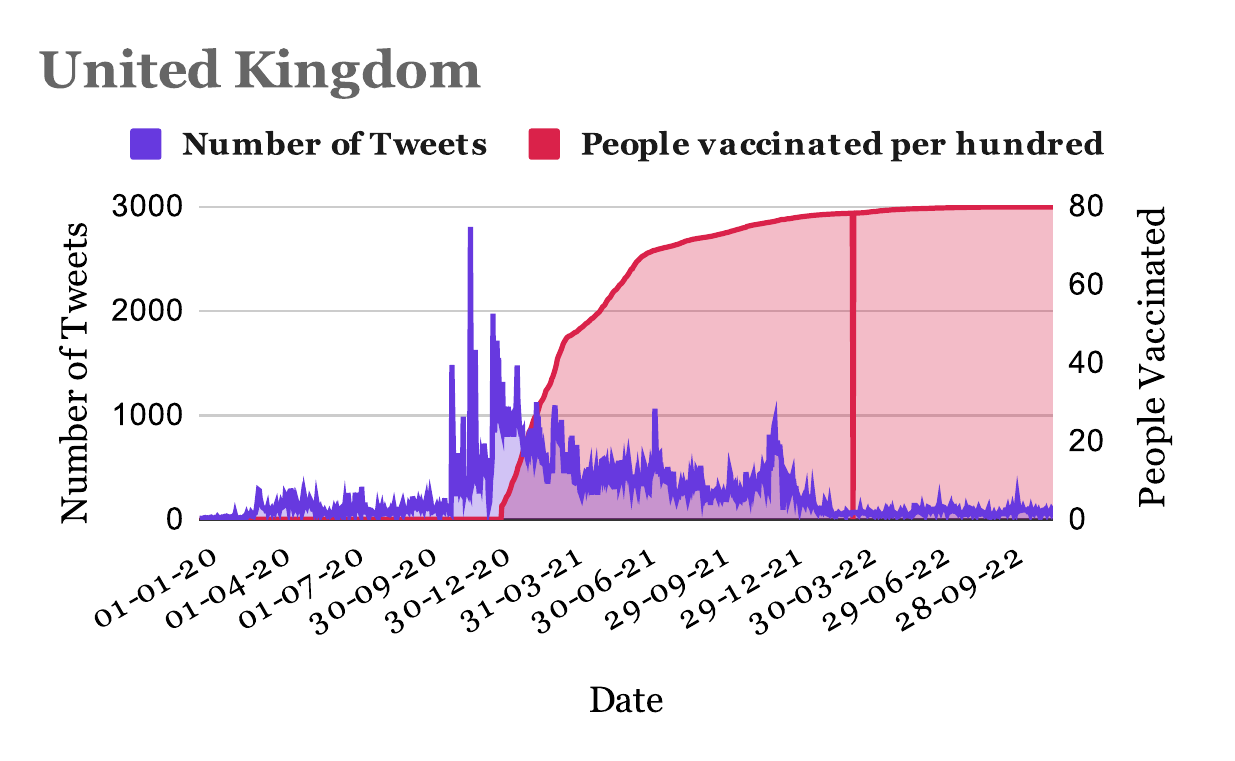}
         \label{fig:2}
     \end{subfigure}\\
     
     \begin{subfigure}[b]{0.49\textwidth}
         \centering
         \includegraphics[width=0.86\textwidth]{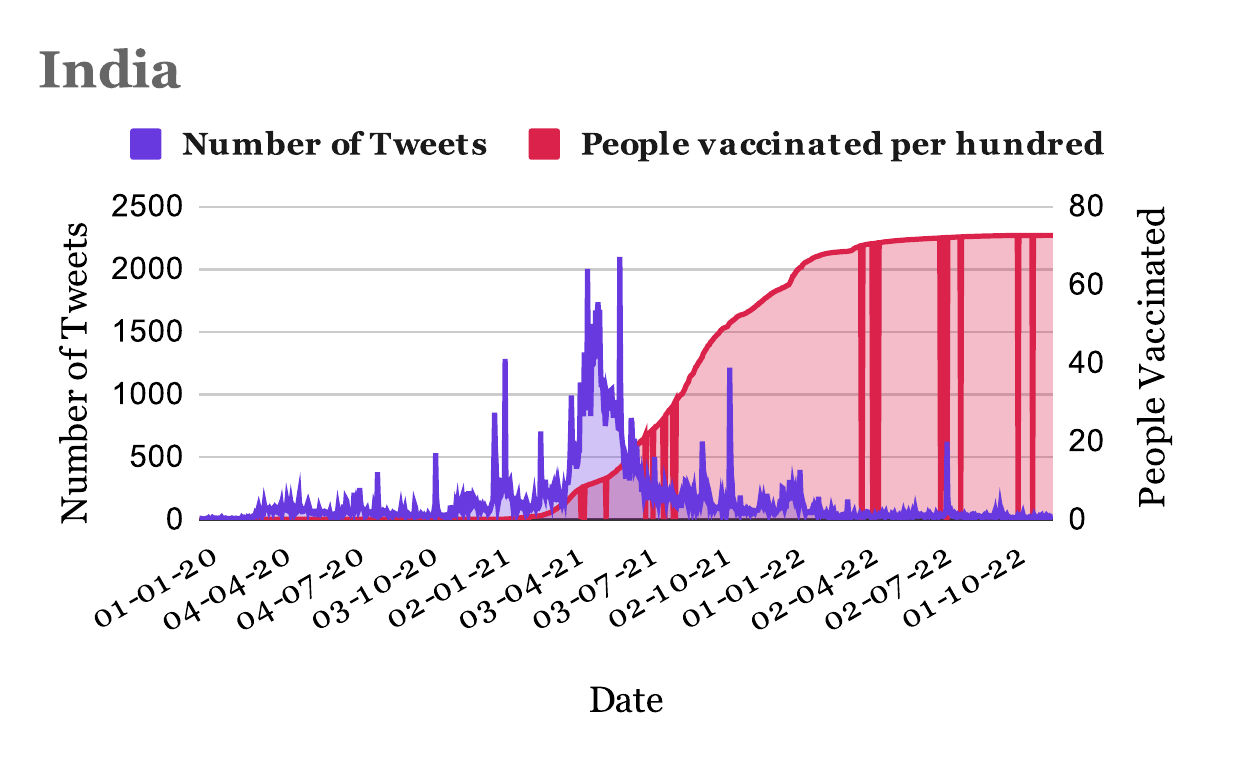}
         \label{fig:3}
     \end{subfigure}
     \hfill
     \begin{subfigure}[b]{0.49\textwidth}
         \centering
         \includegraphics[width=0.86\textwidth]{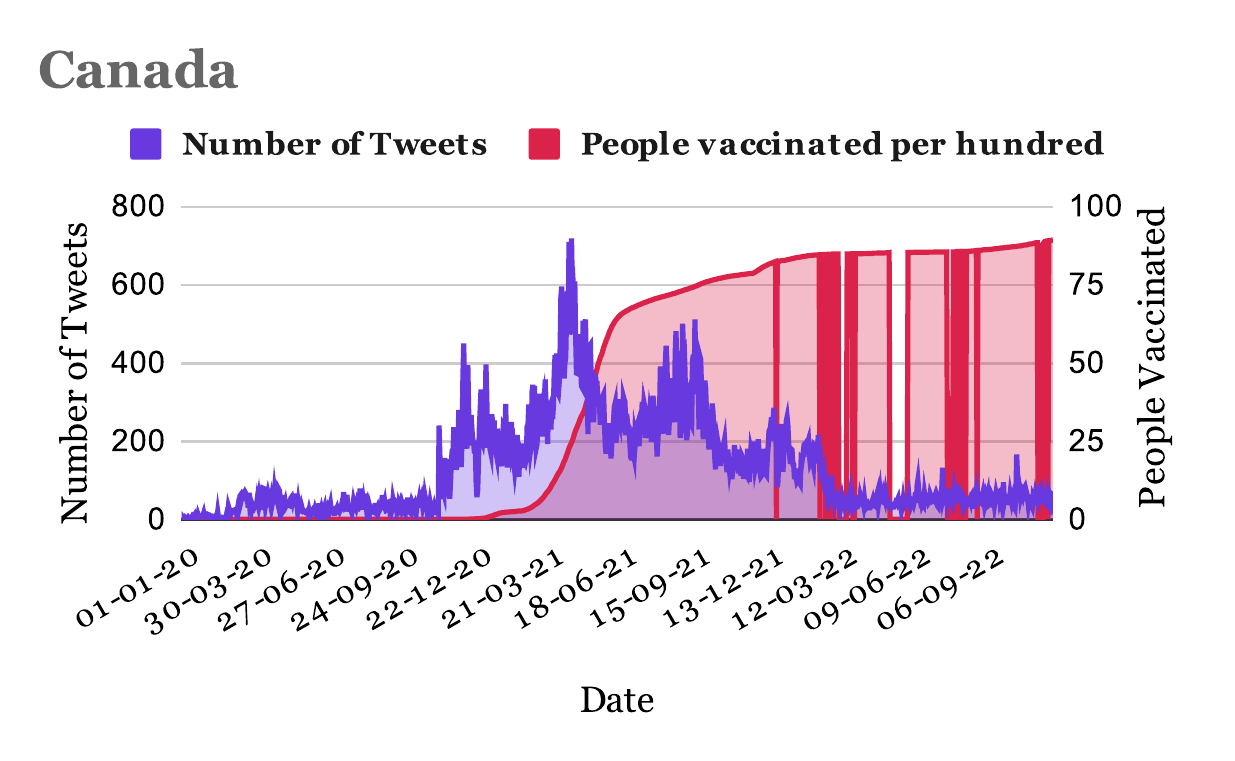}
         \label{fig:4}
     \end{subfigure}\\
     
     \begin{subfigure}[b]{0.49\textwidth}
         \centering
         \includegraphics[width=0.86\textwidth]{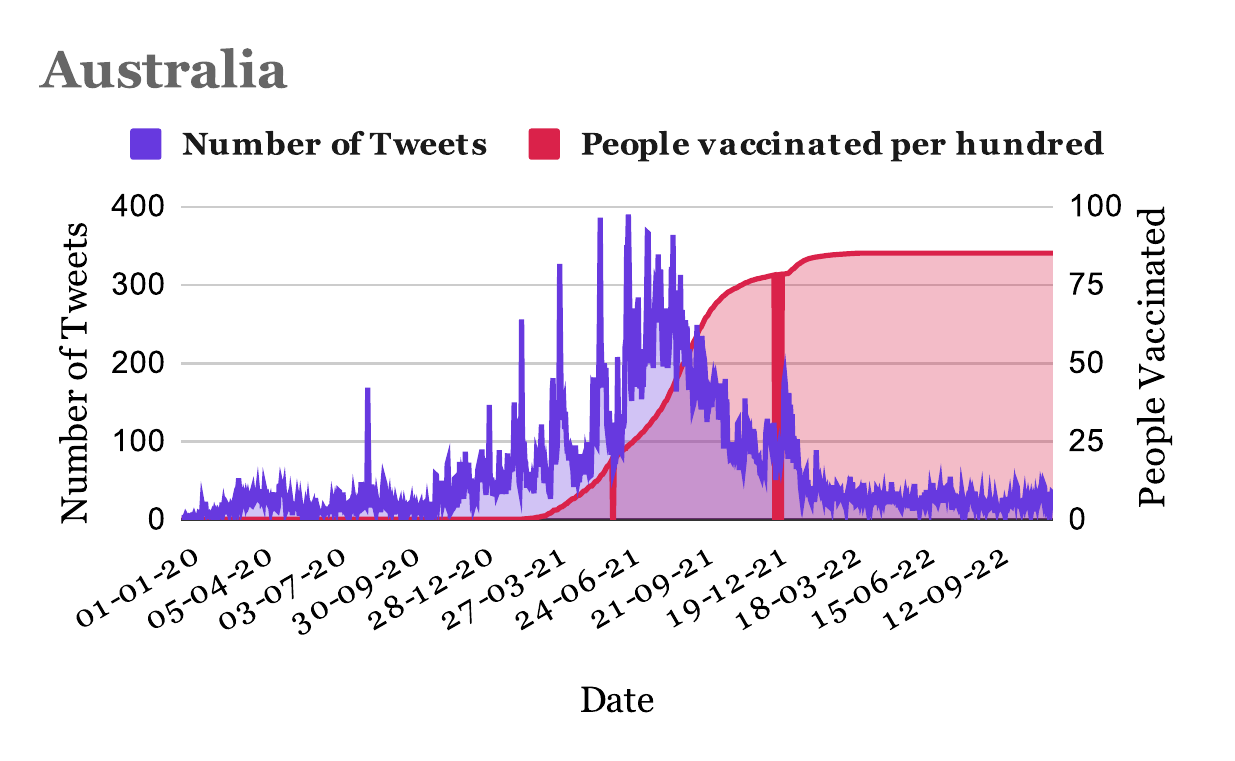}
         \label{fig:5}
     \end{subfigure}
     \hfill
     \begin{subfigure}[b]{0.49\textwidth}
         \centering
         \includegraphics[width=0.86\textwidth]{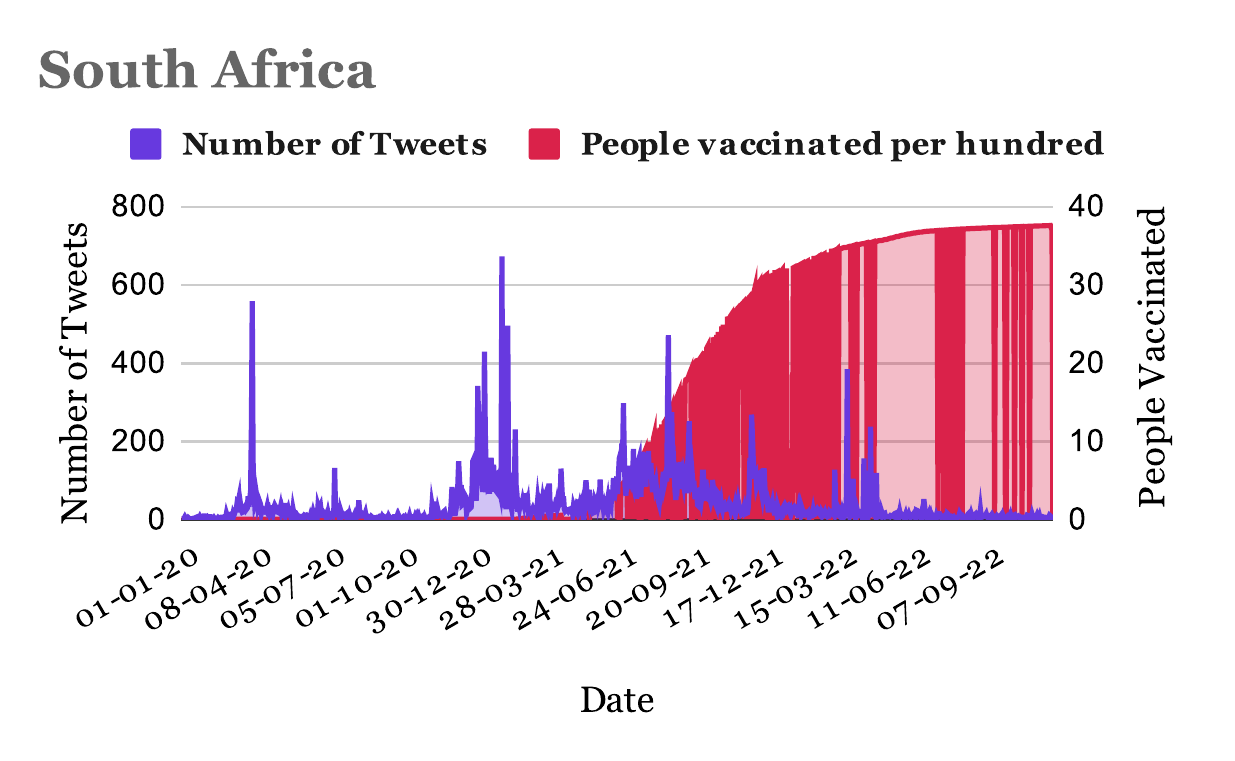}
         \label{fig:6}
     \end{subfigure}
        
\caption{The daily distributions of tweets in the top six countries in the discourse along with their respective cumulative vaccination data. The un-uniformity in the vaccination data for some countries is due to irregular updates.}
\label{six-graphs}
\end{figure*}

The \textit{GeoCovaxTweets} dataset has a total of 1,818,253 geotagged tweets created by 464.9k users between January 01, 2020, and November 25, 2022. Figure \ref{daily-dist} presents the daily distribution of the tweets and the cumulative number of people vaccinated per hundred globally. The source of the vaccination data is \textit{Our World in Data}\footnote{https://ourworldindata.org/}. Out of 1.8 million tweets, 899,592 (49.5\%) are from United States, 263,765 (14.5\%) from United Kingdom, 160,913 (8.8\%) from India, 136,389 (7.5\%) from Canada, and 65,951 (3.6\%) from Australia. In Figure \ref{six-graphs}, we present the daily distribution of tweets in these five countries alongside their respective cumulative vaccination per hundred information. 

Regarding the statewide distribution, England, California, New York, Ontario, and Texas contribute the highest number of tweets. We also explored the different tweet sources in \textit{GeoCovaxTweets}. Twitter's native application for iPhone, Android, and iPad apps are the top 3 sources contributing more than 1.7 million tweets. The list also includes \textit{Instagram} and \textit{dlvr.it} in the top 5 generating above 10k tweets. We provide the top 15 countries, states, and tweet sources with their respective frequencies in Table \ref{top15}. 

We explored the top $n$-grams in the vaccines and vaccination-related discourse. We removed noisy data such as URLs, retweet information, special characters, paragraph breaks, and stop words before extracting bigrams and trigrams. Table 3 list the top 15 bigrams and trigrams in \textit{GeoCovaxTweets}. 
Similarly, we extracted the 15 top-tweeted hashtags, which are as follows: \#covid19, \#vaccine, \#getvaccinated, \#vaccination, \#covid, \#covidvaccine, \#vaccines, \#vaccineswork, \#coronavirus, \#covid\_19, \#vaccinated, \#pfizer, \#wearamask, \#covid19vaccine, and \#astrazeneca.

\begin{table*}[ht]
\caption{The top 15 countries, states, and tweet sources in \textit{GeoCovaxTweets}. Note: $^*$\textit{OpenStreetMap} data returned \texttt{type:state} for these countries.}
\label{top15}
\begin{minipage}{.32\linewidth}
\centering
   \begin{tabular}{c|c}
    \hline
        \textbf{Country} & \textbf{Frequency} \\ 
\hline
United States               & 899,592 \\
United Kingdom              & 263,765 \\
India                       & 160,913 \\
Canada                      & 136,389 \\
Australia                   & 65,951  \\
South Africa                & 40,243  \\
Ireland                     & 31,529  \\
Philippines & 14,992  \\
Malaysia                    & 12,856  \\
New Zealand                 & 11,895  \\
Nigeria                     & 11,711  \\
Pakistan                    & 8,943   \\
Germany                     & 7,910   \\
Kenya                       & 7,853   \\
Spain                       & 6,416   \\
\hline 
\end{tabular}
\end{minipage}
\begin{minipage}{.32\linewidth}
\centering
    \begin{tabular}{c|c}
    \hline
        \textbf{State} & \textbf{Frequency} \\ 
\hline
England$^*$      &     218,477 \\
California &  125,590 \\
New York        & 75,308 \\
Ontario          &  69034\\
Texas           & 68,067 \\
Florida         & 51,433 \\
Illinois        & 31,939 \\
Maharashtra     & 29,976 \\
Ohio            & 29,719 \\
Pennsylvania    & 28,791 \\
Scotland$^*$ & 24,553\\
Washington      & 23,647 \\
Georgia         & 23,558 \\
Massachusetts   & 22,751 \\
Victoria        & 21,933 \\

\hline
\end{tabular}
\end{minipage}
\begin{minipage}{.36\linewidth}
\centering
    \begin{tabular}{c|c}
    \hline
        \textbf{Sources} & \textbf{Frequency} \\ 
\hline
Twitter for iPhone            & 958,579 \\
Twitter for Android           & 776,786 \\
Twitter for iPad              & 31,223  \\
Instagram                     & 16,154  \\
dlvr.it                       & 11,064  \\
Tweetbot for iOS              & 6,450 \\
Twitter Web App               & 5,152   \\
Twitter & 2,020 \\
Twitter for Mac               & 1,806   \\
Foursquare                    & 1,109   \\
Tweetbot for Mac              & 959    \\
Hootsuite Inc.                & 850    \\
TweetCaster for Android       & 664    \\
Foursquare Swarm              & 602    \\
Vaccine Radar                 & 563    \\
\hline
\end{tabular}
\end{minipage}
\end{table*}

\begin{table*}
\caption{The top 15 bigrams and trigrams in \textit{GeoCovaxTweets}.}
\label{}
\begin{minipage}{.5\linewidth}
\centering
    \begin{tabular}{c|c}
    \hline
        \textbf{Bigram} & \textbf{Frequency} \\ \hline
(`covid', `19')          & 121,032    \\
(`covid', `vaccine')     & 72,345     \\
(`get', `vaccine')       & 59,999     \\
(`19', `vaccine')        & 50,897     \\
(`get', `vaccinated')    & 23,756     \\
(`take', `vaccine')      & 21,475     \\
(`getting', `vaccine')   &
19875\\
(`side', `effects')      & 18,847     \\
(`covid19', `vaccine')   & 18,001     \\
(`covid', `vaccines')    & 16,835     \\
(`get', `covid')         & 14,819     \\
(`pfizer', `vaccine')    & 14,262     \\
(`got', `vaccine'),     & 13,911 \\
(`first', `dose')        & 13,392     \\
(`wear', `mask')         & 12,160     \\
 \hline
\end{tabular}
\end{minipage}
\begin{minipage}{.5\linewidth}
\centering
    \begin{tabular}{c|c}
    \hline
        \textbf{Trigram} & \textbf{Frequency} \\ \hline
(`covid', `19', `vaccine')      & 48,653 \\
(`covid', `19', `vaccines')     & 12,428 \\
(`covid', `19', `vaccination')  & 10,905 \\
(`j', `\&', `j')               & 7,671  \\
(`get', `covid', `vaccine')     & 5,243  \\
(`johnson', `\&', `johnson')   & 3,880  \\
(`get', `covid', `19')          & 3,782  \\
(`\&', `j', `vaccine')         & 3,359  \\
(`vaccine', `covid', `19')      & 3,119  \\
(`first', `dose', `vaccine')    & 3,012 \\
(`dose', `covid', `19')         & 2,947  \\
(`dose', `covid', `vaccine')    & 2,852  \\
(`first', `dose', `covid')      & 2,499  \\
(`vaccine', `side', `effects')  & 2,481  \\
(`getting', `covid', `vaccine')       & 2,407  \\
\hline
\end{tabular}
\end{minipage}
\hfill
\end{table*}

\subsection{Dataset Usages} \textit{GeoCovaxTweets} covers a comprehensive conversational dynamics of support, criticism, and hesitance towards the COVID-19 vaccines and vaccination on Twitter. There are several applications and potential usages of \textit{GeoCovaxtweets}. We discuss some of them below:

\begin{itemize}   

\item \textbf{Spatio-temporal analyses:} The dataset can assist in examining how anti-vaccine discussions evolve spatially and tracking temporal relationships between vaccine intakes and social media discussions at the topical- and sentimental-level. The geotagged tweets can be explored to study vaccine hesitancy at the district-, city-, state- or country-level for hot-spot prediction and to get a precise view of how vaccine hesitancy impacts health outcomes. Although we provide state and country information, the dataset also contains tweets with point locations (i.e., geo coordinates) which can be filtered for performing spatial analyses at more granular levels than the state.

\item \textbf{Location inference.} Spatial analyses require geographical data, particularly the origin locations of tweets, rather than the locations mentioned in tweet texts. Dependence on toponyms for spatial analyses can generate biased results due to the \textit{Location A/B problem} \cite{lamsal2022addressing}, as people in location $A$ might participate in the discourse specific to location $B$. Instagram has been reported to generate the most COVID-19-related tweets geotagged with point coordinates, with the majority of them being throwback content, i.e., contexts from the past \cite{lamsal2022did}, concluding that all geotagged tweets from Instagram might not necessarily be originating from the respective geotagged sources. Since, \textit{GeoCovaxTweets} contains both types of geotagged tweets, i.e., point coordinates and place information (bounding box), training models \cite{lamsal2022did} for addressing the Location A/B problem can be another potential use case.

\item \textbf{Downstream datasets:} 
\textit{GeoCovaxTweets} can be used to identify relevant tweets to generate labeled training datasets for fine-tuning language models for downstream tasks.

\item \textbf{Network analysis:} \textit{GeoCovaxTweets} has application in network analysis, particularly for identifying key users or bots and their roles in spreading misinformation, hoaxes, and propaganda concerning COVID-19 vaccines and vaccination.

\item \textbf{Correlation and causality:} Correlation is a measure of computing the relationship between two variables. \textit{GeoCovaxTweets} can be used to perform various types of correlation analysis---such as the correlation between hesitant-type tweets and the number of vaccinations over a period of time. Geotagged tweets have been reported to have variables that granger cause time series such as the daily confirmed COVID-19 cases \cite{lamsal2022twitter}. \textit{GeoCovaxTweets} can be explored if support-, criticize-, and hesitant-type tweets Granger-cause vaccination time series.

\item \textbf{Dashboard:} \textit{GeoCovaxTweets} can also be used in web-based dashboards for visualizing situational data across different geographical contexts \cite{deverna2021covaxxy}.

\end{itemize}

\subsection{Dataset Limitations}
First, Twitter demographics are biased toward the younger generation and tech-aware users. Second, the dataset only includes English-language COVID-19 vaccine discourse missing the opinion of minority dialects and multilingual speakers \cite{jurgens2017incorporating}.
Third, geotagged tweets are required for extracting situational awareness as they involve spatial analyses; however, today less than 1\% of tweets are geotagged \cite{lamsal2021design}. Hence our dataset may not fully reflect all of the Twitter users' conversations around the COVID-19 vaccines and vaccination.

\subsection{Disclaimers}
This dataset should be used only for non-commercial purposes while strictly adhering to Twitter's policies. Note that the number of tweets after hydration can be less than reported, as tweets can be deleted or made private. \textit{GeoCovaxTweets} has not been scrutinized for misinformative and propaganda tweets---the dataset contains all the tweets returned by Twitter's Full-archive search endpoint for applied query+condition. Any location information present in the dataset and the geoinformation generated after hydration should be used while maintaining the privacy of the individuals. The dataset will continue to receive updates until the discourse concerning COVID-19 vaccines and vaccination is relevant. Details regarding updates will be provided on the dataset page.

\section{Conclusion}
This paper presents COVID-19 vaccines and vaccination-specific global geotagged Twitter dataset, \textit{GeoCovaxTweets}, containing more than 1.8 million English language tweets created by 464.9k users from 233 countries and territories between January 01, 2020, and November 25, 2022. The existing datasets either capture the conversational dynamics of the COVID-19 pandemic across every thematic area and therefore include a proportionately lesser number of tweets regarding vaccines and vaccination or are geographically or temporally limited. By releasing \textit{GeoCovaxTweets} to the public, we envision that the dataset will assist in exploring the global spatio-temporal shifts and trends related to the COVID-19 vaccines and vaccination narratives.

\section*{Acknowledgements}
We (the authors) are grateful to \textit{DigitalOcean}\footnote{https://www.digitalocean.com/} for providing the infrastructure needed for this study. The geocoding server was run on a large VM, with 24 VCPUs and 216 gigabytes of memory, running at the University of Melbourne and provided by \textit{Nectar Research Cloud}.

\bibliography{aaai22}

\begin{thebibliography}{23}
\providecommand{\natexlab}[1]{#1}

\bibitem[{Aguas et~al.(2022)Aguas, Gon{\c{c}}alves, Ferreira, and
  Gomes}]{aguas2022herd}
Aguas, R.; Gon{\c{c}}alves, G.; Ferreira, M.~U.; and Gomes, M. G.~M. 2022.
\newblock Herd immunity thresholds for SARS-CoV-2 estimated from unfolding
  epidemics.
\newblock \emph{MedRxiv}, 2020--07.

\bibitem[{Banda et~al.(2021)Banda, Tekumalla, Wang, Yu, Liu, Ding, Artemova,
  Tutubalina, and Chowell}]{banda2021large}
Banda, J.~M.; Tekumalla, R.; Wang, G.; Yu, J.; Liu, T.; Ding, Y.; Artemova, E.;
  Tutubalina, E.; and Chowell, G. 2021.
\newblock A large-scale COVID-19 Twitter chatter dataset for open scientific
  research—an international collaboration.
\newblock \emph{Epidemiologia}, 2(3): 315--324.

\bibitem[{Burki(2019)}]{burki2019vaccine}
Burki, T. 2019.
\newblock Vaccine misinformation and social media.
\newblock \emph{The Lancet Digital Health}, 1(6): e258--e259.

\bibitem[{Chang, Monselise, and Yang(2021)}]{chang2021people}
Chang, C.-H.; Monselise, M.; and Yang, C.~C. 2021.
\newblock What are people concerned about during the pandemic? detecting
  evolving topics about COVID-19 from Twitter.
\newblock \emph{Journal of healthcare informatics research}, 5(1): 70--97.

\bibitem[{Chen et~al.(2020)Chen, Lerman, Ferrara et~al.}]{chen2020tracking}
Chen, E.; Lerman, K.; Ferrara, E.; et~al. 2020.
\newblock Tracking social media discourse about the covid-19 pandemic:
  Development of a public coronavirus twitter data set.
\newblock \emph{JMIR public health and surveillance}, 6(2): e19273.

\bibitem[{DeVerna et~al.(2021)DeVerna, Pierri, Truong, Bollenbacher, Axelrod,
  Loynes, Torres-Lugo, Yang, Menczer, and Bryden}]{deverna2021covaxxy}
DeVerna, M.~R.; Pierri, F.; Truong, B.~T.; Bollenbacher, J.; Axelrod, D.;
  Loynes, N.; Torres-Lugo, C.; Yang, K.-C.; Menczer, F.; and Bryden, J. 2021.
\newblock CoVaxxy: A Collection of English-Language Twitter Posts About
  COVID-19 Vaccines.
\newblock In \emph{ICWSM}, 992--999.

\bibitem[{Hu et~al.(2021)Hu, Wang, Luo, Zhang, Huang, Yan, Liu, Ly, Kacker, She
  et~al.}]{hu2021revealing}
Hu, T.; Wang, S.; Luo, W.; Zhang, M.; Huang, X.; Yan, Y.; Liu, R.; Ly, K.;
  Kacker, V.; She, B.; et~al. 2021.
\newblock Revealing public opinion towards COVID-19 vaccines with Twitter data
  in the United States: spatiotemporal perspective.
\newblock \emph{Journal of Medical Internet Research}, 23(9): e30854.

\bibitem[{Imran, Qazi, and Ofli(2022)}]{imran2022tbcov}
Imran, M.; Qazi, U.; and Ofli, F. 2022.
\newblock Tbcov: two billion multilingual covid-19 tweets with sentiment,
  entity, geo, and gender labels.
\newblock \emph{Data}, 7(1): 8.

\bibitem[{Johnson et~al.(2020)Johnson, Vel{\'a}squez, Restrepo, Leahy, Gabriel,
  El~Oud, Zheng, Manrique, Wuchty, and Lupu}]{johnson2020online}
Johnson, N.~F.; Vel{\'a}squez, N.; Restrepo, N.~J.; Leahy, R.; Gabriel, N.;
  El~Oud, S.; Zheng, M.; Manrique, P.; Wuchty, S.; and Lupu, Y. 2020.
\newblock The online competition between pro-and anti-vaccination views.
\newblock \emph{Nature}, 582(7811): 230--233.

\bibitem[{Jurgens, Tsvetkov, and Jurafsky(2017)}]{jurgens2017incorporating}
Jurgens, D.; Tsvetkov, Y.; and Jurafsky, D. 2017.
\newblock Incorporating dialectal variability for socially equitable language
  identification.
\newblock In \emph{Proceedings of the 55th Annual Meeting of the Association
  for Computational Linguistics (Volume 2: Short Papers)}, 51--57.

\bibitem[{Kata(2012)}]{kata2012anti}
Kata, A. 2012.
\newblock Anti-vaccine activists, Web 2.0, and the postmodern paradigm--An
  overview of tactics and tropes used online by the anti-vaccination movement.
\newblock \emph{Vaccine}, 30(25): 3778--3789.

\bibitem[{KFF(2022)}]{KFF}
KFF. 2022.
\newblock KFF COVID-19 Vaccine Monitor.
\newblock
  \url{https://www.kff.org/coronavirus-covid-19/dashboard/kff-covid-19-vaccine-monitor-dashboard/}.

\bibitem[{Kwok, Vadde, and Wang(2021)}]{kwok2021tweet}
Kwok, S. W.~H.; Vadde, S.~K.; and Wang, G. 2021.
\newblock Tweet topics and sentiments relating to COVID-19 vaccination among
  Australian Twitter users: machine learning analysis.
\newblock \emph{Journal of medical Internet research}, 23(5): e26953.

\bibitem[{Lamsal(2021)}]{lamsal2021design}
Lamsal, R. 2021.
\newblock Design and analysis of a large-scale COVID-19 tweets dataset.
\newblock \emph{applied intelligence}, 51(5): 2790--2804.

\bibitem[{Lamsal, Harwood, and Read(2022{\natexlab{a}})}]{lamsal2022addressing}
Lamsal, R.; Harwood, A.; and Read, M.~R. 2022{\natexlab{a}}.
\newblock Addressing the location A/B problem on Twitter: the next generation
  location inference research.
\newblock In \emph{Proceedings of the 6th ACM SIGSPATIAL International Workshop
  on Location-based Recommendations, Geosocial Networks and Geoadvertising},
  1--4.

\bibitem[{Lamsal, Harwood, and Read(2022{\natexlab{b}})}]{lamsal2022twitter}
Lamsal, R.; Harwood, A.; and Read, M.~R. 2022{\natexlab{b}}.
\newblock Twitter conversations predict the daily confirmed COVID-19 cases.
\newblock \emph{Applied Soft Computing}, 129: 109603.

\bibitem[{Lamsal, Harwood, and Read(2022{\natexlab{c}})}]{lamsal2022did}
Lamsal, R.; Harwood, A.; and Read, M.~R. 2022{\natexlab{c}}.
\newblock Where did you tweet from? Inferring the origin locations of tweets
  based on contextual information.
\newblock \emph{arXiv preprint arXiv:2211.16506}.

\bibitem[{Malagoli et~al.(2021)Malagoli, Stancioli, Ferreira, Vasconcelos,
  Couto~da Silva, and Almeida}]{malagoli2021look}
Malagoli, L.~G.; Stancioli, J.; Ferreira, C.~H.; Vasconcelos, M.; Couto~da
  Silva, A.~P.; and Almeida, J.~M. 2021.
\newblock A look into COVID-19 vaccination debate on Twitter.
\newblock In \emph{13th ACM Web Science Conference 2021}, 225--233.

\bibitem[{Muric et~al.(2021)Muric, Wu, Ferrara et~al.}]{muric2021covid}
Muric, G.; Wu, Y.; Ferrara, E.; et~al. 2021.
\newblock COVID-19 vaccine hesitancy on social media: building a public twitter
  data set of antivaccine content, vaccine misinformation, and conspiracies.
\newblock \emph{JMIR public health and surveillance}, 7(11): e30642.

\bibitem[{Qazi, Imran, and Ofli(2020)}]{qazi2020geocov19}
Qazi, U.; Imran, M.; and Ofli, F. 2020.
\newblock GeoCoV19: a dataset of hundreds of millions of multilingual COVID-19
  tweets with location information.
\newblock \emph{SIGSPATIAL Special}, 12(1): 6--15.

\bibitem[{WHO(2021)}]{WHO}
WHO. 2021.
\newblock COVID-19 coronavirus Vaccines.
\newblock
  \url{https://www.who.int/westernpacific/emergencies/covid-19/covid-19-vaccines}.

\bibitem[{Worldometer(2022)}]{Worldometer}
Worldometer. 2022.
\newblock {COVID-19 coronavirus PANDEMIC}.
\newblock \url{https://www.worldometers.info/coronavirus/}.

\bibitem[{Yang, Hui, and Menczer(2019)}]{yang2019bot}
Yang, K.-C.; Hui, P.-M.; and Menczer, F. 2019.
\newblock Bot electioneering volume: Visualizing social bot activity during
  elections.
\newblock In \emph{Companion Proceedings of The 2019 World Wide Web
  Conference}, 214--217.

\end{thebibliography}

\end{document}